\documentclass[twocolumn]{aastex631}

\usepackage{comment}

\begin{document}

\title{Searching for cosmic vortices}

\correspondingauthor{Marek Niko{\l}ajuk, Tomasz Karpiuk}
\email{m.nikolajuk@uwb.edu.pl, t.karpiuk@uwb.edu.pl}

\author[0000-0003-4075-6745]{Marek Niko{\l}ajuk}
\affiliation{Faculty of Physics, University of Bia{\l}ystok, ul. K. Cio{\l}kowskiego 1L, 15-245 Bia{\l}ystok, Poland}

\author[0000-0001-7194-324X]{Tomasz Karpiuk}
\affiliation{Faculty of Physics, University of Bia{\l}ystok, ul. K. Cio{\l}kowskiego 1L, 15-245 Bia{\l}ystok, Poland}

\author[0000-0002-5273-0641]{Miros{\l}aw Brewczyk}
\affiliation{Faculty of Physics, University of Bia{\l}ystok, ul. K. Cio{\l}kowskiego 1L, 15-245 Bia{\l}ystok, Poland}

\begin{abstract}
Our study focuses on the strong tidal disruption of a cold helium white dwarf passing a black hole. We model the white dwarf as a Bose-Fermi droplet and use quantum hydrodynamic equations to simulate the binary system's evolution. As the white dwarf passes through periastron, it loses a significant amount of mass. This mass falls onto the black hole and forms an accretion disc. Quantized vortices appear in the accretion disc, manifesting as strong electromagnetic radiation signals that exhibit characteristic flickering patterns changing on a timescale of a few seconds. Meanwhile, the white dwarf moves away from the black hole. As the white dwarf moves through space, vortices run along its surface. This elongates its geometry, causing it to rotate and emit gravitational waves. 

\end{abstract}

\keywords{stars: white dwarfs --- stars: black holes --- accretion discs --- hydrodynamics --- methods: numerical}

\section{Introduction} \label{sec:intro}

For a long time, observations of electromagnetic radiation across a wide range of frequencies have provided profound insights into physical processes within and beyond our galaxy and into the primordial universe. Electromagnetic astronomy has revealed a rich diversity of astrophysical and cosmological phenomena by probing different frequency bands. Similarly, observations of gravitational waves across a broad frequency spectrum are expected to offer equally transformative and complementary insights.

Of particular interest are phenomena related to compact binaries involving a black hole (BH) and a white dwarf (WD). Although black hole-white dwarf binaries are difficult to detect, possible formation mechanisms suggest that such systems exist \citep{Ergma98,Tutuk07}. They can be searched in ultracompact X-ray binaries (UCXBs) characterized by a fast orbital period of components around each other and a degenerate donor star (see \citealt{ArmasPadilla23}). Candidates for UCXB consisting stellar-mass BH have already been discovered (e.g. 47 Tuc, \citealt{MillerJones15}; IGR J17285-2922, \citealt{Stoop21}, RZ 2109, \citealt{Dage24}).

Interestingly, a search of X-ray data from the nearby galaxies NGC 4636 and NGC 5128 uncovered two sources of ultraluminous X-ray flares not associated with galactic centers \citep{Irwin16}. The first source flared once with an estimated peak X-ray luminosity of $9 \times 10^{40}$ erg s$^{-1}$, while the second flared five times and was about an order of magnitude weaker. \cite{Sivakoff05} reported another similar X-ray source in the elliptical galaxy NGC 4697 that exhibited two flares, and more recently \cite{Tiengo22} described recurrent flaring activity from the ultraluminous X-ray source XMMU J122939.7+075333 located in a globular cluster in the galaxy NGC 4472. The properties of these sources are difficult to reconcile with those of well-known Galactic accreting X-ray sources, such as type I and II bursts observed in some low-mass X-ray binaries or beamed emission from stellar-mass black holes. Consequently, it has been proposed that they may represent a new class of fast transients \citep{Irwin16}. One possible explanation is that the observed X-ray bursts originate from tidal stripping of a white dwarf orbiting an intermediate-mass black hole \citep{Krolik11,Shen19,Karpiuk21,Tiengo22,Nikolajuk25}. More recently, two studies \citep{Miniutti19,Arcodia21} reported sequences of quasi-periodic, highly energetic X-ray eruptions, possibly driven by the disruption of a compact object orbiting a supermassive black hole \citep{Arcodia21,Liu23}.

Close encounters between white dwarfs and black holes are also expected to be relatively common. Such encounters lead to the development of large accretion discs surrounding the black holes. In this study, we show that these discs exhibit a flickering behavior, which is associated with the emergence of quantized macroscopic vortices, or cosmic vortices, penetrating the accretion disc.


The detection of gravitational waves (GWs) by the LIGO and Virgo laser interferometer experiments \citep{Abbott16} has opened an entirely new observational window into the cosmos, with far-reaching consequences for fundamental physics, astrophysics, and cosmology. At present, ground-based detectors such as LIGO, Virgo, and KAGRA are sensitive to GW signals at frequencies near $10^{2}\,$Hz. In contrast, the upcoming Laser Interferometer Space Antenna (LISA) will be optimized for frequencies around $10^{-2}\,$Hz. Consequently, the intermediate frequency range near $1\,$Hz remains largely uncharted.


Atom interferometers, which exploit the wave nature of atoms, are anticipated to close the sensitivity gap between terrestrial and space-based interferometric detectors \citep{Baynham25}. This expectation aligns with the results presented here. Our simulations indicate that a white dwarf escaping from a black hole produces gravitational waves at frequencies of about $1\,$Hz. These gravitational waves arise because cosmic vortices propagate along the surface of the escaping white dwarf, effectively rotating it.

\section{Modeling formation of cosmic vortices} \label{sec:modeling}

Helium white dwarfs may not solidify as they cool down \citep{Gabadadze08b}. Although low temperatures allow for the recombination of helium nuclei and electrons, the system should be considered a Bose-Fermi mixture rather than an atomic gas. This is due to the extremely high particle densities within a white dwarf, which lead to an extremely small average inter-particle distance. This distance is two orders of magnitude smaller than the Bohr radius, and recombined atoms strongly overlap. Therefore, the helium white dwarf should be described as a two-component system, even at very low temperatures. Even at the moment they are formed, the temperatures of white dwarfs are a few orders of magnitude lower than the Fermi temperature for electrons \citep{ST83}. Therefore, the fermionic component can be safely regarded as a zero-temperature gas. On the other hand, at low temperatures, the helium nuclei may form a charged condensate, as discussed in Refs. \cite{Gabadadze08a, Gabadadze08b, Gabadadze09, Mosquera10}. This would justify treating a cold helium white dwarf as a Bose-Fermi droplet \citep{Rakshit18}.

In this paper, we study the tidal disruption of a cold helium white dwarf passing a black hole. If the white dwarf is significantly disturbed, it releases a substantial amount of mass onto the black hole, forming an accretion disc. Then, it moves away. We observe the formation of cosmic vortices in the accretion disc, which contribute to the flickering phenomenon. Cosmic vortices also accompany a white dwarf traveling through space, causing gravitational wave emission.

To model tidal disruption of a cold helium white dwarf passing a black hole, we employ the formalism of quantum hydrodynamics (see, for example, \cite{Madelung,Frolich,Wong,Wheeler,MarchDeb}). The hydrodynamic equations for bosons and fermions can be expressed as Schr\"odinger-like equations using the inverse Madelung transformation (see, for example, \cite{Dey98,Domps98,Grochowski17,Grochowski20,Karpiuk21}). For fermions, this is merely a mathematical transformation that replaces the density and velocity (assumed to be irrotational) fields used in a hydrodynamic description with a single complex function. In order to obtain a stable, self-bound solution of the hydrodynamic equations in the form of a Bose-Fermi droplet, quantum corrections due to quantum fluctuations must be included (see \cite{Rakshit18}).

Next, the Bose-Fermi droplet is placed in the gravitational field of an artificial black hole, modeled as a non-rotating black hole described by the Schwarzschild metric. To simplify the analysis, we employ the well-established approximation introduced by \cite{Paczynsky80} for the potential energy of a test particle moving in Schwarzschild space-time. Within this approach, the radial potential is replaced by the pseudo-Newtonian form $V_{PN}(r)=-GM_{BH}/(r-R_S)$, where $R_S=2GM_{BH}/c^2$ denotes the Schwarzschild radius. This potential accurately reproduces the location of the innermost stable circular orbits. Because $V_{PN} (r)$ is independent of angular momentum, the motion of a test particle can be treated as that of a particle moving in three-dimensional space under the influence of the pseudo-Newtonian potential. Consequently, the equations of motion for a Bose-Fermi droplet orbiting a fixed black hole can be written as
\begin{eqnarray}
&& i \hbar \frac{\partial \psi_B}{\partial t} = (H^{eff}_B + V_{PN}\,m_B)\,\psi_B  \nonumber \\
&& i \hbar \frac{\partial \psi_F}{\partial t} = (H^{eff}_F + V_{PN}\,m_F)\,\psi_F   \,,
\label{eqmWDBH}
\end{eqnarray}  
where the complex fields $\psi_B({\bf r},t)$ and $\psi_F({\bf r},t)$ represent the bosonic wave function and the fermionic pseudo-wave function, respectively. The effective nonlinear single-particle Hamiltonians $H^{eff}_B$ and $H^{eff}_F$ are detailed in the Appendix \ref{hydeq} (see also \cite{Nikolajuk25}).

\section{Flickering of the accretion disc}  \label{sec:ad}

We solve Eq. (\ref{eqmWDBH}) numerically using the split-operator technique \citep{Gawryluk17,Swislocki26} for the trajectories of an atomic white dwarf exhibiting strong tidal disruption. We consider a Bose-Fermi droplet consisting of bosonic ${}^{133}$Cs and fermionic ${}^{6}$Li atoms. Such mixtures are being studied experimentally nowadays \citep{Weidemuller14,Chin17,Chin18,Lippi25}. The droplet contains $N_B=1460$ bosonic and $N_F=100$ fermionic atoms. Initially, it is located three times its radius away from the artificial black hole, which is more than $1000\, R_S$. The initial velocity of the droplet is chosen so that the white dwarf passes the black hole while shedding a large portion of its mass (about $60\,\%$). Then, an accretion disc forms, and finally, the white dwarf moves away.

\begin{figure}[t]
\includegraphics[width=9.0cm]{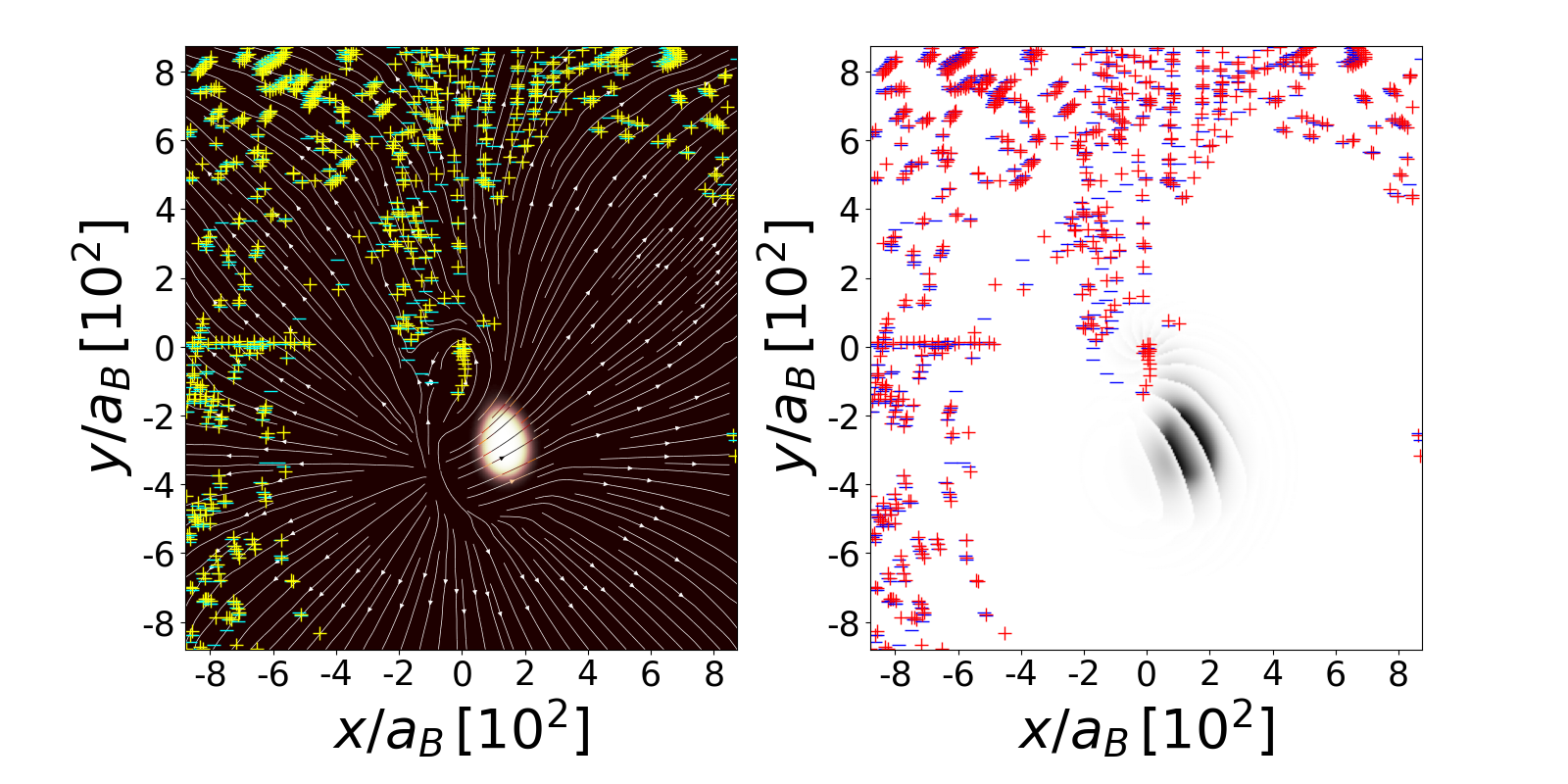}  \\   \vspace{0.4cm}
\includegraphics[width=9.0cm]{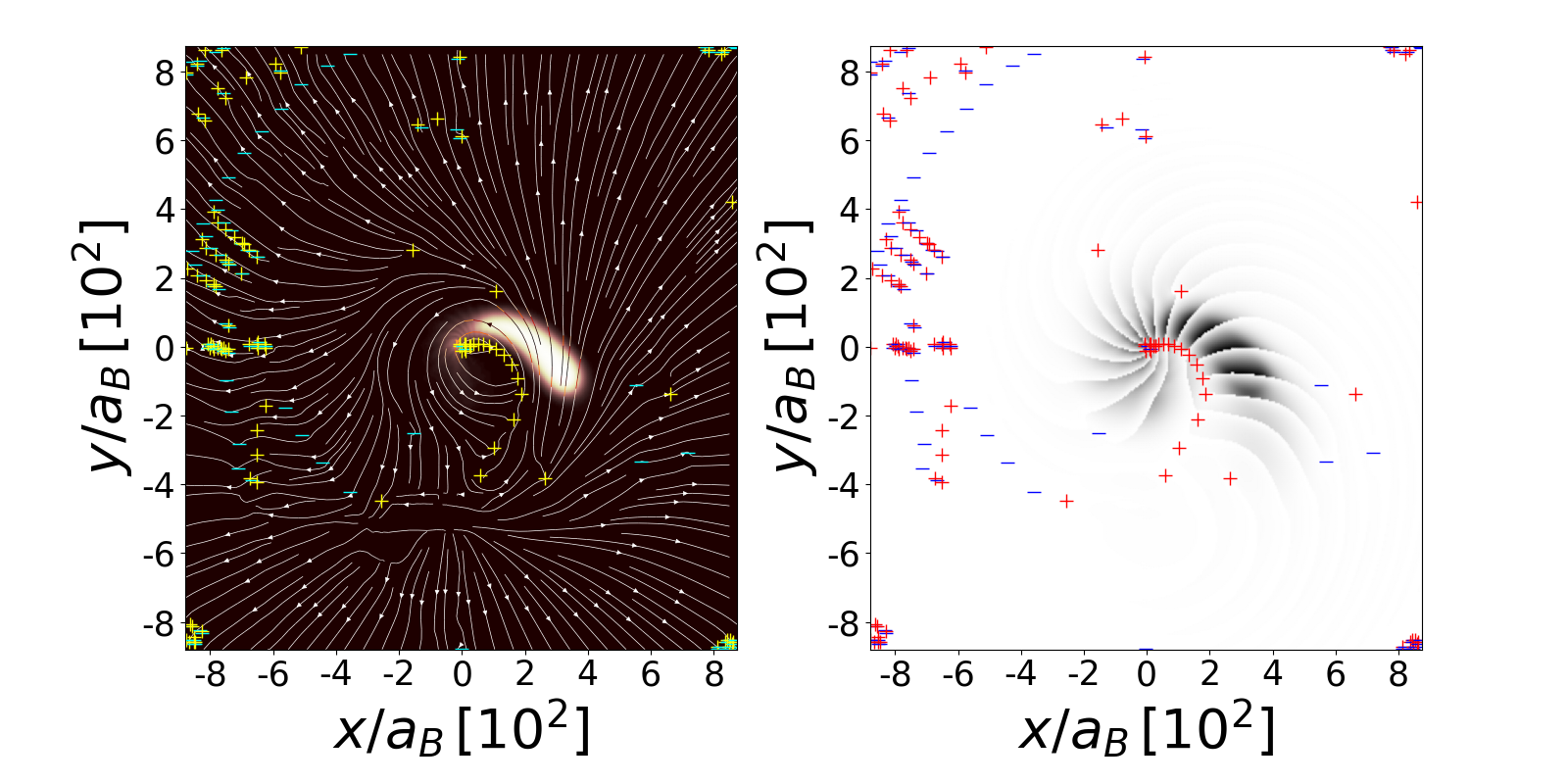}  \\   \vspace{0.4cm}
\includegraphics[width=9.0cm]{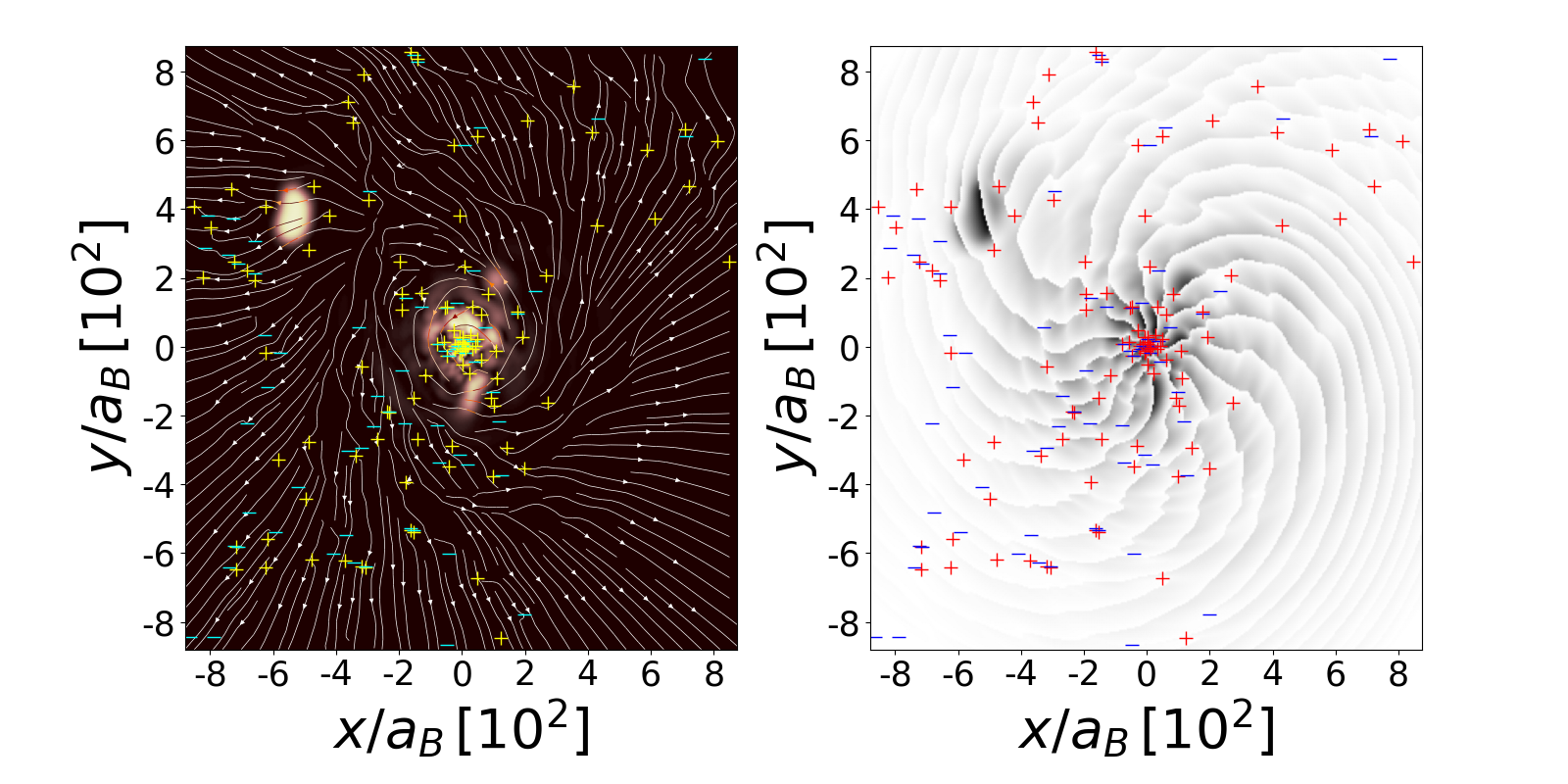}
\caption{Bosonic density (left column) and quantum phase (right column) at the $z=0$ plane for different time instants. The density and time are in units of $1/a_B^3$ and $m_B a_B^2/\hbar$, respectively. The positions of the positive and negative quantized vortices are marked by red/yellow plus signs and blue minus signs, respectively. Initially, there are no vortices associated with the white dwarf (top frame), although many are visible in the background noise. Later on, the white dwarf drags some vortices into an accretion disc that is forming (middle frame). Finally (bottom frame), the white dwarf moves away from a black hole, taking a few vortices with it.  }
\label{vorticesini}
\end{figure}

A movie showing the tidal disruption of a white dwarf passing through periastron is available at \cite{movie26vortices}. Before approaching periastron, the white dwarf, which is initially spherically symmetric, becomes deformed (see Fig.~\ref{vorticesini}, top frame). When the tidal disruption event begins, the mass falls onto the black hole, creating an accretion disc. Figure \ref{vorticesini} (middle frame) shows that the falling mass drags positive, quantized vortices into the accretion disc. Note that an escaping white dwarf also takes quantized vortices with it (bottom frame).

\begin{figure}[htb]
\includegraphics[width=8.2cm]{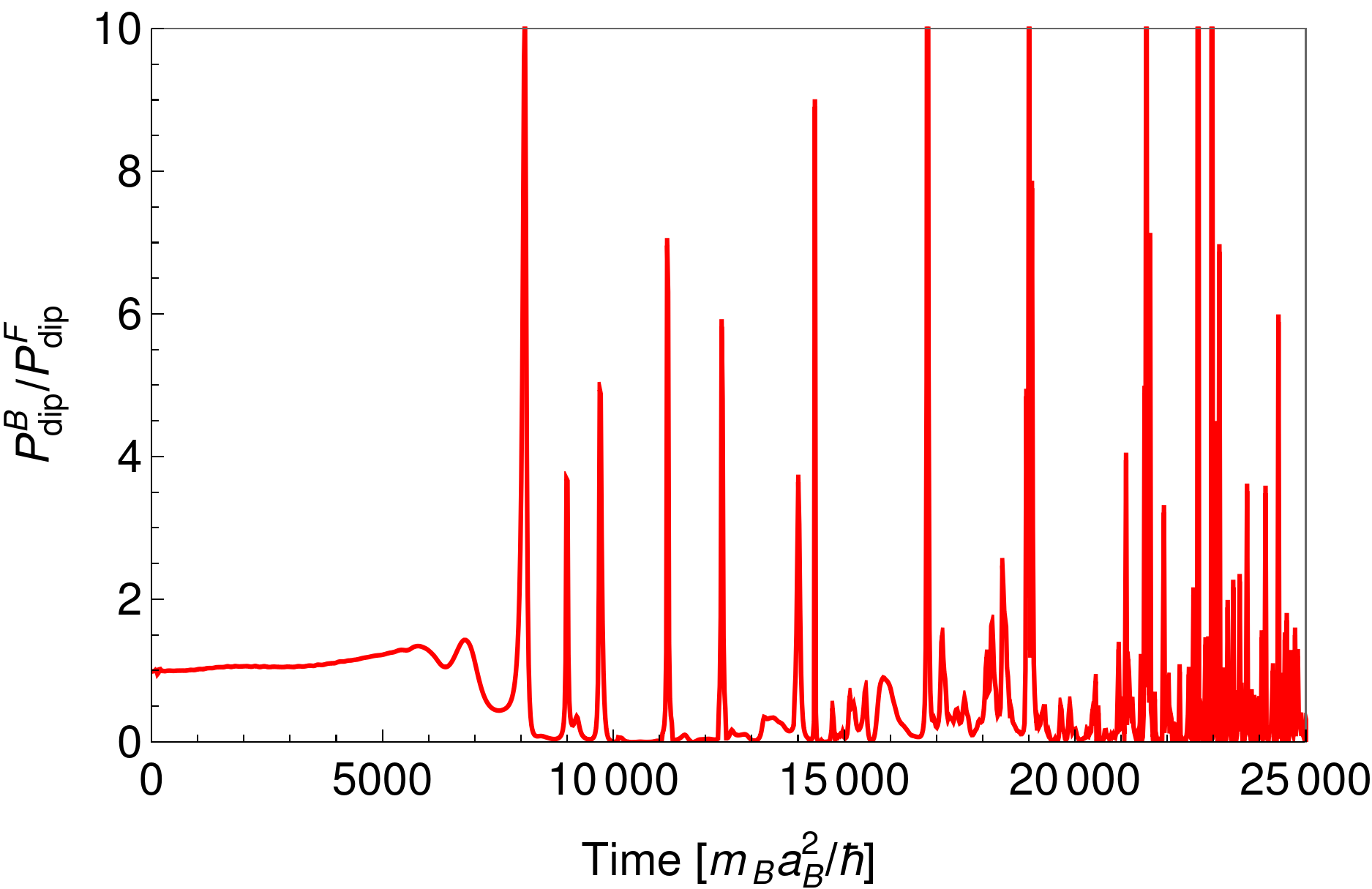} 
\caption{$P_{dip}^B/P_{dip}^F$ as a function of time. Several peaks are visible within the time interval between $5000$ and $10000$, indicating the preliminary stage of cosmic vortex formation in the accretion disc. The dipole electromagnetic radiation power is calculated within a cylinder with a radius of $500\, a_B$. Later, the dynamics of the accretion disc become more complex.}
\label{pbpfdip}
\end{figure}

In this section, we focus on the electromagnetic radiation properties of the accretion disc. We will discuss the dynamics of escaping white dwarf in the next section. A video at \cite{movie26vortices} illustrates the intricate dynamics of an accretion disc, leading to the production of intense electromagnetic radiation. As we will argue below, these dynamics and the resulting radiation blasts are somehow correlated with the appearance of quantized vortices. Assuming the fermionic density follows the bosonic one, we have $\bar{n}_F({\bf r},t)=\bar{n}_B({\bf r},t)$, where $\bar{n}_{B(F)}({\bf r})$ means atomic densities normalized to one (see \cite{Nikolajuk25}). The contributions to the electromagnetic dipole radiation coming from the fermionic and bosonic components are given by $P_{dip}^{B(F)} \sim |\ddot{{\bf p}}_{B(F)}|^2$, where the electric dipole moment is
\begin{eqnarray}
{\bf{p}}_{B(F)} = q_B N_B  \int  {\bf{r}}\, \bar{n}_{B(F)} ({\bf r},t)\,  d^3 r 
\end{eqnarray}
and $q_B$ is the charge of the bosonic particle. Both contributions can now be compared. In the above case, one has $P_{dip}^B/P_{dip}^F=1$. This is true, but only during the initial stages of evolution (see Fig.~\ref{pbpfdip}). However, the process begins later on, when vortices enter and start to penetrate the accretion disc. But vortices appear only in a bosonic component, leaving only a density dip in the fermionic counterpart. Since charged particles in a vortex undergo additional rotational motion, the electromagnetic radiation coming from the spatial region accompanied by vortices must be strengthened. In this case, the $P^B_{dip}$ signal is obviously stronger than the $P^F_{dip}$ signal. Indeed, several well-defined spikes are visible in Fig.~\ref{pbpfdip} after time $5000$, when the first vortices enter the accretion disc and the initial matter falling around the black hole hits the spout of matter that is still strengthening the accretion disc (see the movie at \cite{movie26vortices}). Later, however, a multitude of peaks appear in the spectrum, clearly demonstrating the complexity of the accretion disc's dynamics.

Further studies of the dynamics of an accretion disc can be performed using statistical analysis. To determine if dipole and quadrupole radiation is correlated with vortices entering the accretion disc, we calculate the covariance:
\begin{eqnarray}
\rm{cov}(X,Y) = \langle X Y \rangle - \langle X \rangle \langle Y \rangle  \,,
\label{covariance}
\end{eqnarray}  
where random variable $X$ represents the temporal change of dipole or quadrupole power originating from a small spatial plaquette, while the variable $Y$ indicates whether or not there is a vortex in the plaquette. The entire plane in which the white dwarf moves is divided into an array of spatial plaquettes. Each plaquette accumulates radiation from the entire spatial region above and below it, perpendicular to the plane of motion. The random variable $Y$ takes the values $+1$ ($-1$) for a positively (negatively) charged quantized vortex or zero if no vortex is present in the plaquette. The averaging $\langle \cdot \rangle$ is done over a specific time period, usually a few time steps.

\begin{figure}[t]
\includegraphics[width=8.6cm]{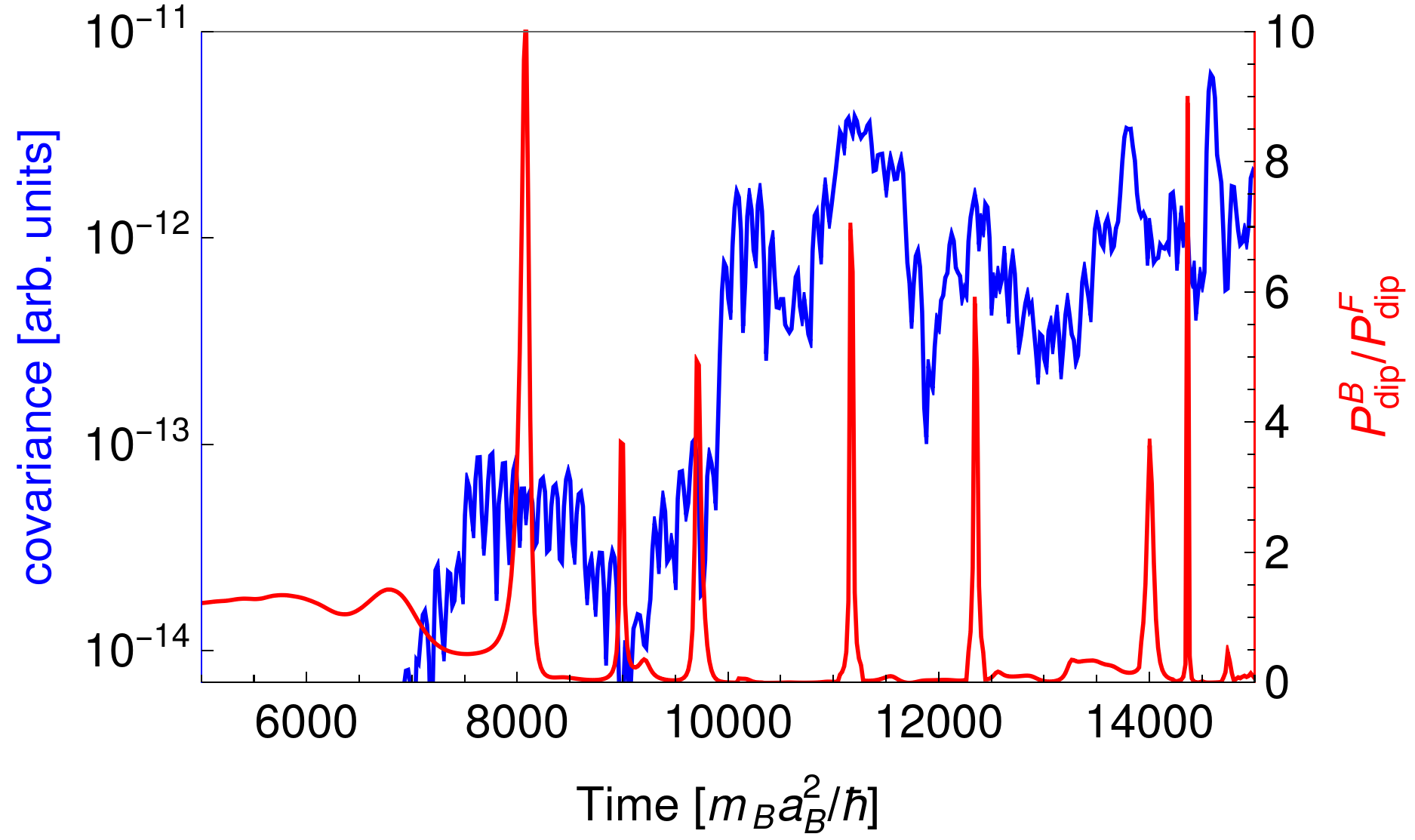} 
\caption{Ratio $P_{dip}^B/P_{dip}^F$ (in red) and the maximum covariance (in blue) are plotted as a function of time. The red spikes representing the ratio clearly correlate with time intervals of high covariance.  }
\label{pbpfcov}
\end{figure}

In Fig.~\ref{pbpfcov}, we plot the covariance (in blue) and the ratio $P_{dip}^B/P_{dip}^F$ (in red) up to $15000$ code units. The maximum value of the modulus of covariance over the grid of spatial plaquettes is plotted in blue. Most of the spikes in the ratio $P_{dip}^B/P_{dip}^F$ are correlated with time intervals of high covariance. This suggests that vortices may be responsible for triggering strong electromagnetic radiation.

\begin{figure}[hbt]
\includegraphics[width=8.8cm]{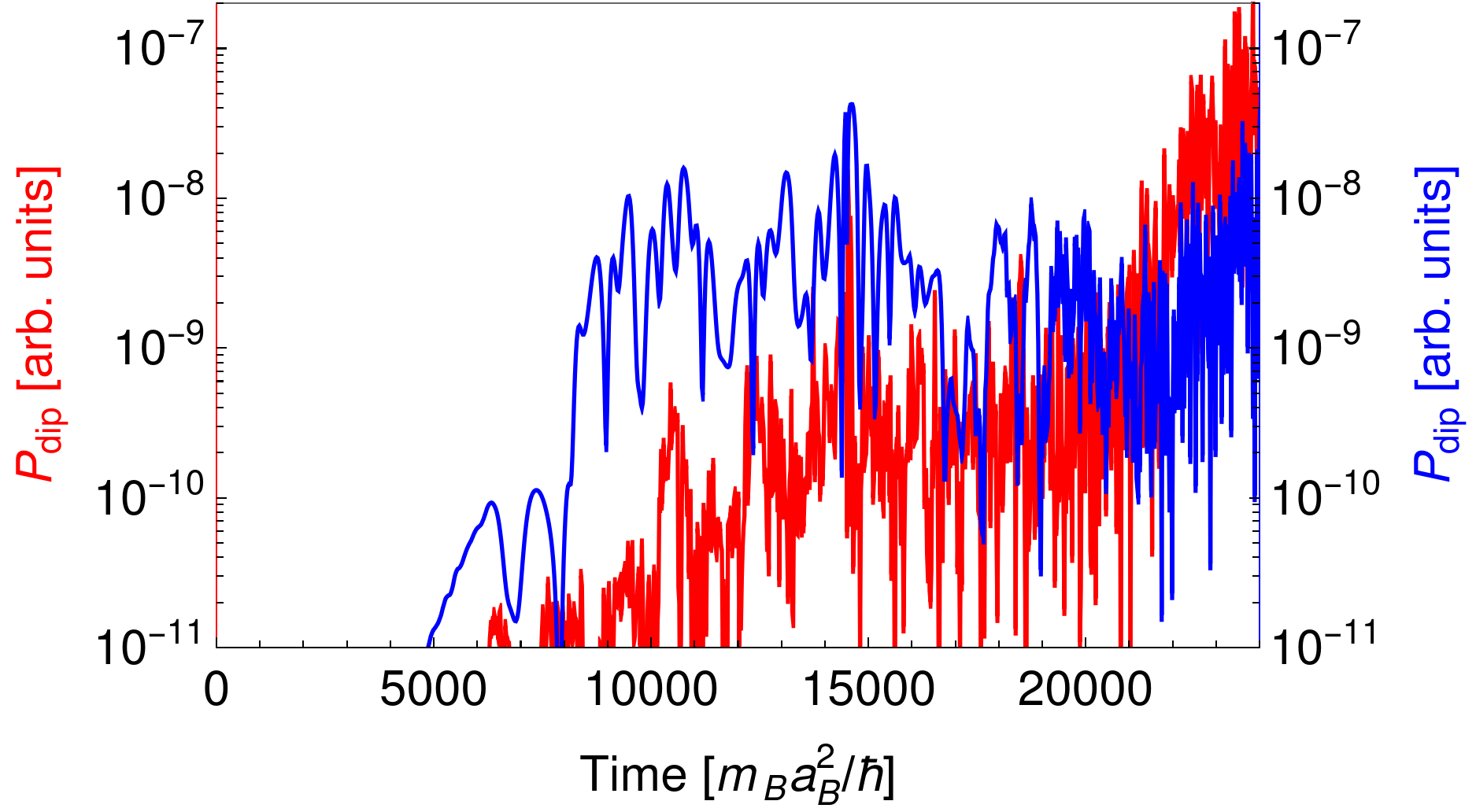} 
\caption{Electromagnetic dipole radiation power coming from a region with a radius of $700\, a_B$ (blue curve) and originating from the ten most powerful plaquettes with vortices (red curve) as a function of time. }
\label{dipvortex}
\end{figure}

To strengthen the role of vortices in generating electromagnetic radiation further, we calculate the power of dipole radiation originating from the ten most energetic plaquettes containing vortices. The results are shown in Fig. \ref{dipvortex} as a function of time (red curve) and are compared with the power emitted by the entire accretion disc (blue curve). Initially, between times $5000$ and $16000$, the radiation is generated mainly by accelerating charged matter (blue curve), the emission by the region with vortices is an order of magnitude lower. However, this interplay changes after time $16000$, when the white dwarf is absent and the accretion disc is left alone. After $20000$, the radiation from the region with vortices becomes dominant.

\begin{figure}[htb]
\includegraphics[width=8.6cm]{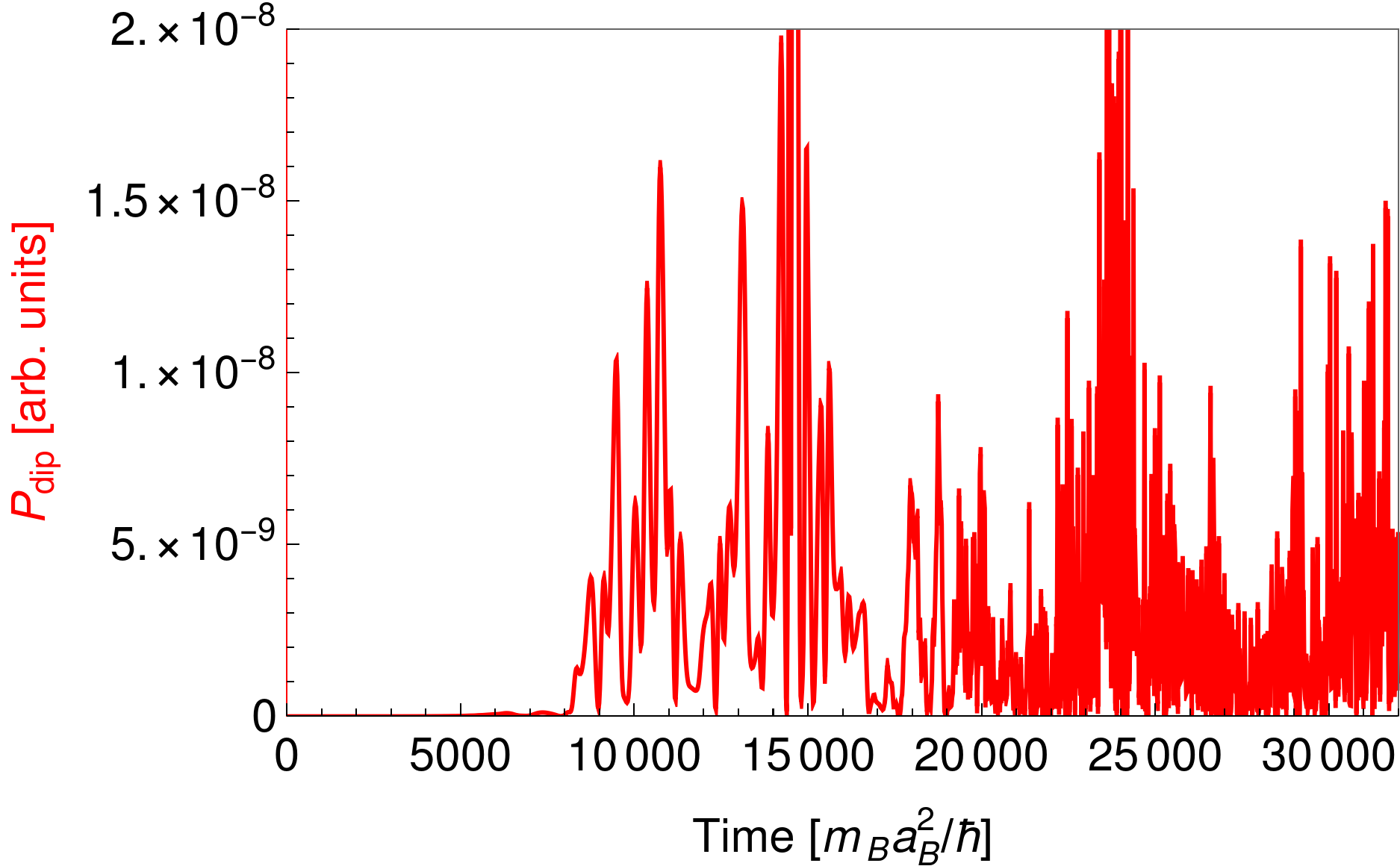} 
\caption{Electromagnetic dipole radiation coming from an accretion disc, when the white dwarf has already departed (times after $16000$) and during the preformation period (times between $5000$ and $16000$). Aperiodic behavior, i.e., flickering, is expected and confirmed by an analysis of the power spectrum density (see Fig.~\ref{flickering}).}
\label{disc}
\end{figure}

\begin{figure}[thb]
\includegraphics[width=8.6cm]{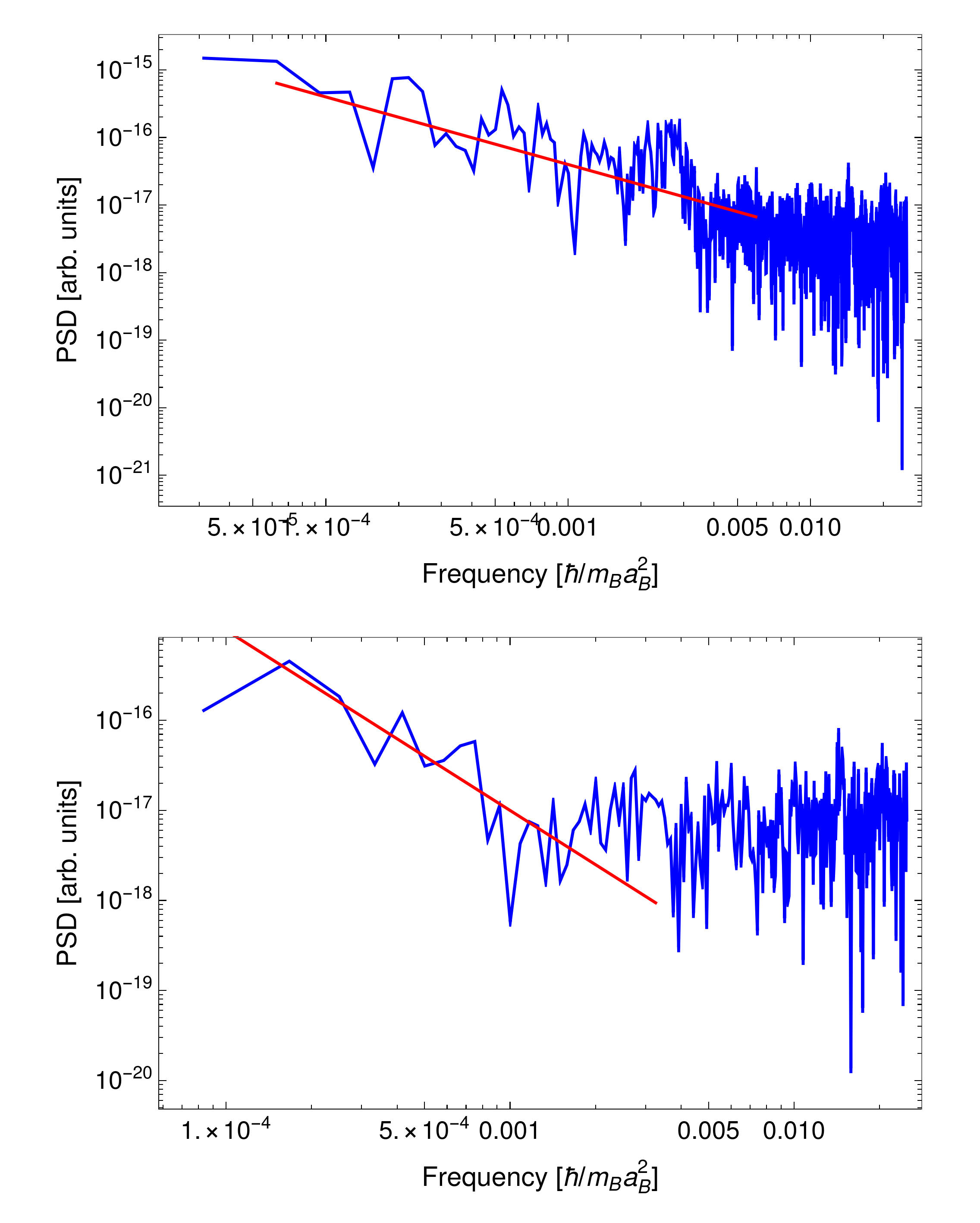} 
\caption{Power spectrum density of the radiation coming from the accretion disc is shown as a function of frequency ($f$). $PSD\sim1/f$ (solid red line) exhibits flicker noise when the preformation period is included in the analysis, followed by a white noise tail at the highest frequencies (top frame). However, when examining radiation from a well-formed accretion disc, the $PSD$ changes qualitatively to $PSD\sim1/f^2$ (solid red line), again with a lingering white noise tail (bottom frame). }
\label{flickering}
\end{figure}

Finally, we analyze the electromagnetic radiation of an accretion disc on its own, i.e., after $16000$ code units, when the white dwarf has already escaped (see the movie at \cite{movie26vortices}). Fig. \ref{disc} shows the power of electromagnetic dipole radiation as a function of time, including the preformation period (the time between $5000$ and $16000$). There are clearly no regularities, and the degree of irregularity increases after the formation of the accretion disc. The behavior of the power spectral density (PSD), defined as the modulus square of the Fourier transform of the dipole power, $|{\cal{F}}[P_{dip}]|^2$, supports this observation. As shown in Fig.~\ref{flickering}, the power spectral density (solid red line) undergoes a qualitative change as the accretion disc forms. This behavior is related to the formation of vortices in the accretion disc, the number of which stabilizes once the white dwarf goes away.

In astrophysics, power spectral density is an important characteristic that allows us to understand the physical properties of the objects we are studying. For instance, we can deduce the mass of a black hole in an active galactic nucleus (AGN) from its PSD. A mature, steady-state AGN does not exhibit a single slope in its PSD. The PSD's behavior follows a bending power law model, where the slope transitions smoothly at a specific ''bending frequency''. At low frequencies, the PSD shows $1/f$ dependence, and beyond the bending frequency, the slope changes to $1/f^2$ 
\citep[e.g.,][]{Uttley02,Markowitz03,McHardy06,GMV12}. The transition point between the $1/f$ and $1/f^2$ slopes is governed by the system's physical scale, which allows us to determine the black hole's mass at the heart of AGNs \citep[see][]{bcz01,Niko04,GPP21}.

Fig.~\ref{flickering} shows that the power spectral density of an accretion disc changes from $PSD\sim1/f$ during the birth phase to $PSD\sim1/f^2$ during the mature phase. This kind of evolutionary path, which differs from the AGN's behavior, has been observed in real-world astrophysical systems \citep{Chakra26}. When a wandering star passes too close to a supermassive black hole, it is violently ripped apart by tidal forces. The left debris fall toward the black hole, creating a highly chaotic, shock-dominated, newly born accretion disc. During this youthful phase, the broadband noise of the light curve shows a power-law index (approximately $1/f^{\gamma}$, with $\gamma \approx 0.5$), followed by a transitions into a flat, Poisson white noise floor at higher frequencies. This birth phase lasts from minutes to hours. After this period, the system settles into a stable accretion disc with corona. The PSD tilts and depends on frequency, approximately following a $1/f^{\gamma}$, with $\gamma \approx 1.0$, power law, followed again by a white noise tail.

In our study of the tidal disruption of a cold white dwarf by a black hole, the PSD dependence on density behaves similarly, but the birth phase lasts only a few seconds. Additionally, the accretion disc persists after the white dwarf escapes. Thus, the system discussed here differs from short gamma-ray burst events, in which the accretion disc is formed and disappears within a few seconds. Therefore, it can be classified as a distinct type of astrophysical system based on its structure and dynamics.

As the white dwarf undergoes periastron passage, tidal stripping tears away mass. This causes the falling matter to fragment into distinct blocks due to nonlinear quantum effects. This chaotic, fragmented mass injection generates multiscale fluctuations and triggers the nucleation of numerous giant quantized vortices in the bosonic component of the disc. This active, turbulent quantum vortex network is the literal source of $1/f$ flicker noise. Once the white dwarf escapes, the mass injection ceases. The newly formed disc is then allowed to stabilize. Without a continuous supply of fragmented matter to drive the turbulence, the number of quantized vortices settles. The system shifts to a localized, memoryless quantum diffusion or relaxation process, which turns the PSD into a steep $1/f^2$ profile. This observation could be considered evidence of cosmic vortices in the accretion disc.

\section{Gravitational waves from an escaping white dwarf}   \label{sec:wd}

In this section, we focus on the properties of the white dwarf moving away from the black hole. The white dwarf's motion for the considered orbit is shown in the movie available at \cite{movie26vortices}, as discussed in the previous section. Before periastron is approached, the white dwarf becomes deformed (see Fig.~\ref{vorticesini}, top frame). When the tidal disruption event begins, the mass falls onto the black hole and creates an accretion disc. Figure \ref{vorticesini} (middle frame) shows that the falling mass drags positive, quantized vortices into the accretion disc. This leads to blasts of electromagnetic radiation (Sec.~\ref{sec:ad}).
 
At $t \approx 16000$ in units of $m_B a_B^2/\hbar$, the white dwarf begins to depart from the black hole, see the movie at \cite{movie26escapingWD}. The movie clearly demonstrates that the white dwarf's geometry is not axially symmetric with respect to the axis perpendicular to the plane of motion. The movie shows the white dwarf rotating around its center of mass. Its rotation period is $T_{rot} \approx 9200$. Scaling this result to the realm of astronomical objects in the manner proposed in \cite{Nikolajuk25} means that the white dwarf rotates with a period of $T_{rot} \approx 5\,$s ($\omega_{rot} \approx 2\pi \times 0.2\,$Hz), assuming an initial helium white dwarf with a mass of $m_{wd}=0.4 M_{\odot}$ falls onto a stellar-mass black hole with a mass of $M_{\rm BH} = 5\, M_{\odot}$. This number is obtained by multiplying the numerical value by a scaling factor of $\mathcal{T}$ (see Appendix \ref{scaling}).

\begin{figure}[hbt]
\includegraphics[width=8.0cm]{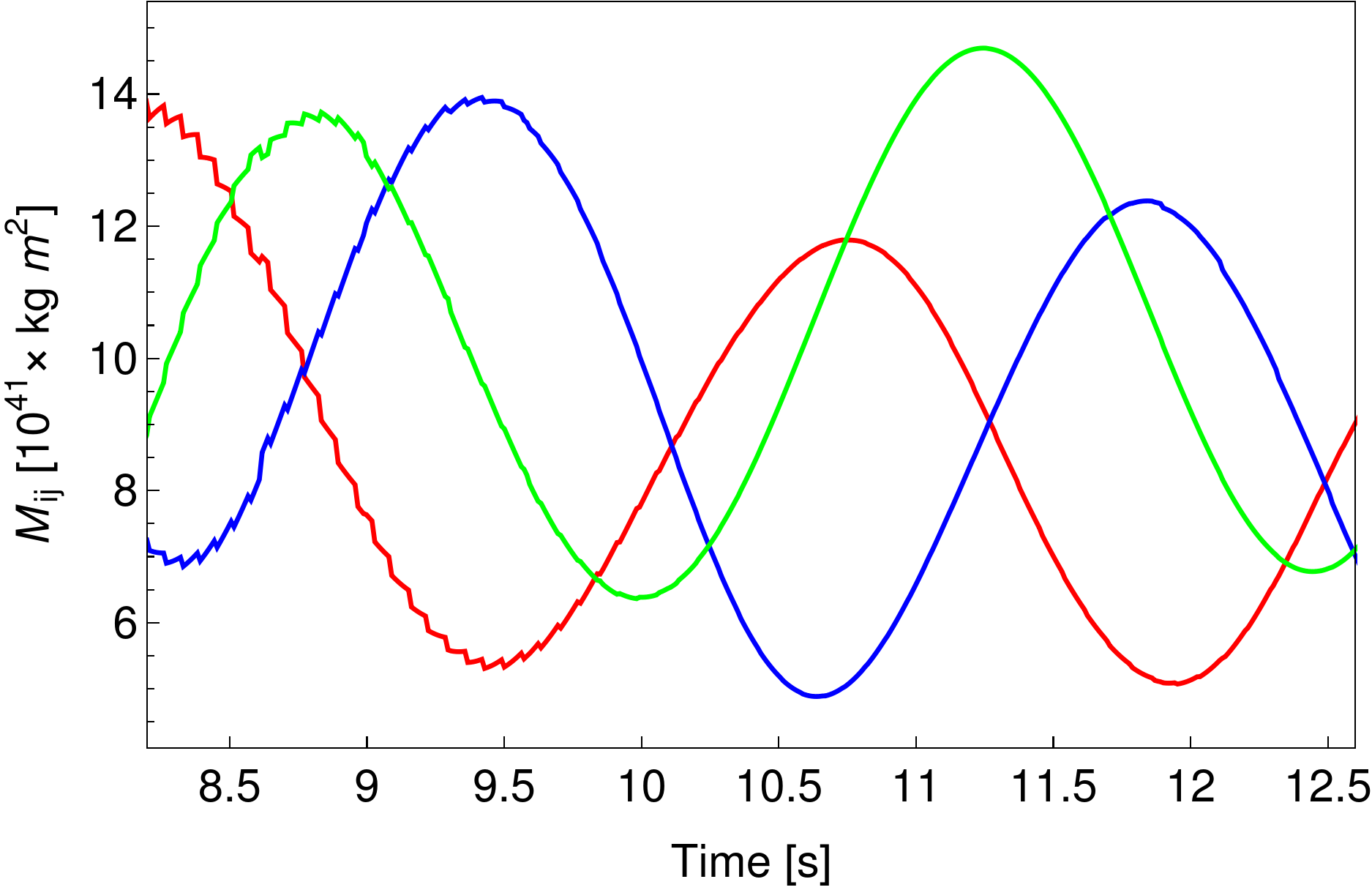}
\caption{Second moments of the mass density ($M_{ij}$, with $i,j=x,y$) of an escaping white dwarf are shown as a function of time. Only times after the white dwarf has clearly departed the black hole and accretion disc are considered. The curves depict $M_{xx}$ (red), $M_{yy}$ (blue), and $M_{xy}$ (green) moments. Note that the $M_{xy}$ moment has been shifted so that it overlaps with two other moments. }
\label{wdescaping}
\end{figure}

Clearly, a rotating white dwarf traveling through space (see movie at Ref. \cite{movie26escapingWD}) is a source of gravitational radiation, as is an anisotropic rotating rigid body \citep{Maggiore}. In the quadrupole approximation the gravitational wave amplitude depends on the second moment of the mass density, $M_{ij}=\int \rho({\bf{r}})\, r_i\, r_j\, d^3 r$, where $\rho({\bf{r}})$ is the mass density. The second moments of the mass density of a rigid body rotating around one of its principal axes (here, collinear with the fixed $z$ axis) change over time as (up to irrelevant constants)
\begin{eqnarray}
M_{xx} &=& -\frac{I_1 - I_2}{2} \cos{(2 \omega_{rot} t)}   \nonumber  \\
M_{xy} &=& -\frac{I_1 - I_2}{2} \sin{(2 \omega_{rot} t)}   \nonumber  \\
M_{yy} &=& +\frac{I_1 - I_2}{2} \cos{(2 \omega_{rot} t)}  \,.
\label{quadrupolemoments}
\end{eqnarray}
Here, $I_1$ and $I_2$ are the moments of inertia of a rigid body with respect to principal axes lying in the plane of motion and $\omega_{rot}$ is the rotation frequency around the axis perpendicular to this plane. In Fig. \ref{wdescaping} we plot second moments of the mass density $M_{ij}$ ($i,j=x,y$) of an escaping white dwarf, calculated with respect to its mass center, as a function of time. An agreement with the formulas (\ref{quadrupolemoments}) is remarkable even though some modulation of amplitudes is observed. The amplitudes the $M_{ij}$ moments change in time give $I_1 - I_2 \approx 7\times 10^{41}\,$kg m$^2$ and the numerical values of the moments are scaled to the realm of astronomical objects according to the prescription presented in Appendix \ref{scaling}. Then the amplitudes of the gravitational waves for plus and cross polarizations, when observing the binary system from the above, are \citep{Maggiore}
\begin{eqnarray}
h_{+} &=& h_0 \cos{(2\pi f_{gw} t)}  \nonumber  \\
h_{\times} &=& h_0 \sin{(2\pi f_{gw} t)}  \,,
\label{gwamplitudes}
\end{eqnarray}
where
\begin{eqnarray}
h_0 &=& \frac{1}{r}\, \frac{4\pi^2 G}{c^4} (I_1 - I_2)\, f_{gw}^2  \,,
\end{eqnarray}
$r$ is the distance from the system, and $f_{gw}=2\, \omega_{rot}/2\pi$ is the frequency of the gravitational waves. Taking the numerical results characteristic of a rotating white dwarf into account, one can rewrite the gravitational wave amplitude as follows: 
\begin{eqnarray}
h_0 &\approx&  1.1\times 10^{-21}\,  \left(\frac{I_1 - I_2}{10^{42}\, {\rm kg}\, {\rm m}^2}\right) \left(\frac{10\, {\rm kpc}}{r}\right)\, \left(\frac{f_{gw}}{1\, {\rm Hz}}\right)^{\!2}   \,.  \nonumber  \\
\label{strain}
\end{eqnarray}
Then, using a typical galactic distance of $r=10$ kpc (i.e., assuming a white dwarf is expelled by a black hole in the Milky Way), one could observe gravitational waves from a rotating white dwarf with a strain of $\sim 10^{-22}$.

The rotational energy of escaping white dwarf decreases in time, because of emission of gravitational waves, as \citep{Maggiore}
\begin{eqnarray}
\frac{d E_{rot}}{d t} = - \frac{32 G}{5 c^5}\, (I_1 - I_2)^2\, \omega_{rot}^6   \,.
\label{rotenergy}
\end{eqnarray}
In our case, the right-hand side of Eq. (\ref{rotenergy}) is estimated as $\sim  10^{32}\,$J/s. Note that the rotational energy, $E_{rot}=I_3\, \omega_{rot}^2 /2$ (with $I_3\approx 6.6\times 10^{41}\,$kg m$^2$), itself is of the order of $10^{41}\,$J. Then, assuming the emission of the gravitational waves is the only mechanism for losing rotational energy one has
\begin{eqnarray}
\frac{d\, \omega_{rot}}{d t} = - \frac{32 G}{5 c^5}\, \frac{(I_1 - I_2)^2}{I_3}\, \omega_{rot}^5   \,.
\label{rotloss}
\end{eqnarray}
In our case it is $\sim 10^{-10}\,$s$^{-2}$, which is extremely slow rate.

Roughly speaking, the frequency of the gravitational wave emitted by the white dwarf is given by $\omega_{gw}=2\, \omega_{rot} \approx 2\pi \times 0.4\,$Hz. Currently operating terrestrial laser interferometer detectors (LIGO, Virgo, and KAGRA) are sensitive to the gravitational waves at much higher frequencies, around $10^2\,$Hz. Conversely, the Laser Interferometer Space Antenna detector will be sensitive to frequencies around $10^{-2}\,$Hz. However, atom interferometers \citep{Baynham25} are planned to detect waves with frequencies of about $1\,$Hz and will be able to detect gravitational waves originating from a rotating white dwarf traveling through space.


The prolate shape of the white dwarf departing the black hole is an implication of the fact that the white dwarf takes away the quantized vortices. If the angular momentum of the bosonic superfluid per particle is not a multiple of Planck's constant divided by $2\pi$ (i.e., multiple of $\hbar$), the rotating superfluid may become unstable \citep{Butts99}. Consequently, the vortices run along the surface of the system with a small angular momentum per particle -- about $0.3\, \hbar$ in our case -- causing the system to lose its axial symmetry. For a white dwarf leaving a black hole, this results in the emission of gravitational waves. Thermal effects typically destroy persistent currents, so they presumably limit the timescale for producing gravitational radiation.

\section{Conclusions}

In summary, our analysis suggests the existence of a new type of astronomical object, which we refer to as a "cosmic vortex". These vortices may form when a cold helium white dwarf passes close to a black hole and deposits a significant amount of mass onto it. The resulting accretion disc contains numerous vortices that emit strong electromagnetic signals when they come into contact with surrounding matter. The power spectral density of the emitted radiation exhibits a characteristic flickering pattern that changes over time as the system transforms from its initial phase to maturity. This transition occurs within a few seconds, setting the system apart from others that are already known. Eventually, the white dwarf moves away from the black hole, carrying some of the vortices with it. A white dwarf that travels through space while rotating due to the motion of vortices along its surface is a source of gravitational waves.

\begin{acknowledgments}
Part of the results were obtained using computers at the Computer Center of the University of Bialystok.
\end{acknowledgments}

\appendix

\section{Hydrodynamic equations}
\label{hydeq}
The equations of motion for the Bose-Fermi droplet moving in the field of a fixed black hole are given by Eq. (\ref{eqmWDBH}), where the effective nonlinear single-particle Hamiltonians $H^{eff}_B$ and $H^{eff}_F$ have the form \citep{Karpiuk20}
\begin{eqnarray}
H^{eff}_B &=& -\frac{\hbar^2}{2 m_B}\nabla^2 
 + g_B\, |\psi_B|^2 + \frac{5}{2} C_{LHY}\, |\psi_B|^3    \nonumber \\
&+&  g_{BF}\, |\psi_F|^2 + C_{BF}\, |\psi_F|^{8/3} A(w,\alpha)   \nonumber \\
&+&  C_{BF}\,|\psi_B|^2 |\psi_F|^{8/3}\, \frac{\partial A}{\partial \alpha} \frac{\partial \alpha}{\partial n_B}   
\label{HamBF1}
\end{eqnarray}    
and  
\begin{eqnarray}  
H^{eff}_F &=& -\frac{\hbar^2}{2 m_F}\nabla^2 
+ (1 - \xi) \frac{\hbar^2}{2 m_F} \frac{\nabla^2 |\psi_F|}{|\psi_F|} \nonumber \\
&+& \frac{5}{3} \kappa_k |\psi_F|^{4/3} + g_{BF}\, |\psi_B|^2  \nonumber \\
&+& \frac{4}{3} C_{BF}\, |\psi_B|^2 |\psi_F|^{2/3} A(w,\alpha)   \nonumber \\
&+&  C_{BF}\,|\psi_B|^2 |\psi_F|^{8/3}\, \frac{\partial A}{\partial \alpha} \frac{\partial \alpha}{\partial n_F}  \,.
\label{HamBF}
\end{eqnarray}  
The bosonic wave function and the fermionic pseudo-wave function are normalized to the total number of particles in bosonic and fermionic components, $N_{B(F)} = \int d\mathbf{r}\, |\psi_{B(F)}|^2$, respectively. Other parameters are $\kappa_k = (3/10)\,(6\pi^2)^{2/3}\,\hbar^2/m_F$, $\xi=1/9$, $C_{LHY}=64/(15\sqrt{\pi})\,g_B\, a_B^{3/2}$, and $C_{BF}=(6 \pi^2)^{2/3} \hbar^2 a_{BF}^2 / 2 m_F$. Moreover \citep{Giorgini02}
\begin{eqnarray}
A(w,\alpha) = \frac{2(1+w)}{3w}\left(\frac{6}{\pi}\right)^{2/3}\int^{\infty}_0 {\rm d}k \int^{+1}_{-1}{\rm d}{\Omega}
\left[ 1 -\frac{3k^2(1+w)}{\sqrt{k^2+\alpha}}
\int^{1}_0{\rm d}q q^2 \frac{1-\Theta(1-\sqrt{q^2+k^2+2kq\Omega})}{\sqrt{k^2+\alpha}+wk+2qw\Omega}  \right], \nonumber\\
\label{A}
\end{eqnarray}
where $w=m_B/m_F$, $\alpha=2w (g_B n_B/\varepsilon_F)$, $\varepsilon_F=(5/3) \kappa_k n_F^{2/3}$, and $\Theta()$ being the step theta-function.

\section{Scaling}
\label{scaling}

All numerical results from the simulations are scaled to the realm of astronomical objects. The size and mass of the Bose-Fermi droplet used in the simulations can easily be scaled. This is done as follows:
\begin{equation}
\begin{array}{ccc}
r_{\rm wd} \  &   = & \mathcal{A} \,\, r_{\rm num} \ [a_B] \\ 
m_{\rm wd} \  &   = & \phantom{ai}\mathcal{B} \,\, m_{\rm num} \ [m_B]   \,,
\label{sizemass}
\end{array}
\end{equation}
where $r_{wd}=0.0155 R_{\odot}$ and $m_{wd}=0.4 M_{\odot}$ are an exemplary helium white dwarf radius and mass (see \cite{Althaus10,Corsico19,Saumon22}). Since the droplet radius is $r_{\rm num} \simeq 1\mu$m ($=200 a_B$) and the droplet mass is $m_{\rm num} = 1460\, m_B + 100\, m_F = 3.23 \times 10^{-22}$ kg, the dimensionless parameters $\mathcal{A}$ and $\mathcal{B}$ are: $\mathcal{A}\approx 10^{13}$ and $\mathcal{B}\approx 2\times 10^{51}$.
The way time is scaled,
\begin{equation}
\begin{array}{ccc}
t_{\rm real}  \  &   = & \mathcal{T}\, t_{\rm num} \ [m_B a_B^2/\hbar]  \,,
\label{time}
\end{array}
\end{equation}
can be determined by comparing the total energies of a Bose-Fermi droplet and a white dwarf \citep{Nikolajuk25} or by analyzing the timescales of a stripping process in a binary system \citep{Shen19}. In both cases $\mathcal{T} \sim 10^4$.

In order to study the gravitational waves emitted by an escaping white dwarf, we must scale the inertia tensor of the Bose-Fermi droplet. This tensor is defined as follows:
\begin{eqnarray}
I_{ij} &=& \int  (r^2 \delta_{ij} - r_i r_j)\, 
\Big( m_B n_B ({\bf{r}}) + m_F n_F ({\bf{r}}) \Big)   \,  d^3 r   \nonumber \\
&=&  m_B N_B\, \bar{I}_{ij}   \,,
\label{inertiatensor}
\end{eqnarray}
with $m_B$ and $m_F$ being the masses of bosonic and fermionic component particles, ($i,j=x,y,z$), and
\begin{eqnarray}
\bar{I}_{ij} &=& \int  (r^2 \delta_{ij} - r_i r_j)\, 
\Big( \bar{n}_B ({\bf{r}}) + \frac{m_F N_F}{m_B N_B}\, \bar{n}_F ({\bf{r}}) \Big)   \,  d^3 r  \nonumber  \\
&\approx&  \int  (r^2 \delta_{ij} - r_i r_j)\, \bar{n}_B ({\bf{r}}) \,  d^3 r   \,.
\label{inertiatensor1}
\end{eqnarray}
The last term in the upper row of Eq. (\ref{inertiatensor1}) can be safely neglected since the mass of the bosonic component dominates in the cesium-lithium atomic Bose-Fermi droplets assumed in our simulations, as well as in white dwarf stars. The inertia tensor, given by Eq. (\ref{inertiatensor}), scales from the numerical to the astrophysical realm as
\begin{eqnarray}
(I_{ij})_{\rm astro} &=&  \Big( m_B N_B\, \bar{I}_{ij} \Big)_{\rm astro} 
= (m_B N_B)_{\rm astro}\, \Big( \bar{I}_{ij} \Big)_{\rm astro}   \nonumber \\
&=& m_{wd}\,\, \mathcal{A}^2 \, \Big( \bar{I}_{ij} \Big|_{num}\, [a_B^2] \Big) \,,  \nonumber \\
\label{scalinginertiatensor}
\end{eqnarray}
where $m_{wd}$ is the mass of the white dwarf and the numerical value of the tensor, $\bar{I}_{ij}$, is taken from numerical simulations. The scaling factor, $A^2$, appears due to the units of $\bar{I}_{ij}$. Note that the second moments of the mass density, $M_{ij}=\int \rho({\bf{r}})\, r_i r_j\, d^3 r$, where $\rho({\bf{r}})$ is the density, also scale according to Eq. (\ref{scalinginertiatensor}).

\bibliography{BHWD}{}
\bibliographystyle{aasjournal}

\end{document}